\documentclass[smallextended]{svjour3}       
\smartqed  
\usepackage{graphicx}
\usepackage{marvosym}
\begin{document}

\title{A M\"ossbauer study of the magneto-structural coupling effect in SrFe$_2$As$_2$ and SrFeAsF\thanks{This work was supported by the National Natural Science Foundation of China under Grants No. 10975066.}
}


\author{Zhiwei Li \and Xiaoming Ma \and Hua Pang \and Fashen Li 
}


\institute{Zhiwei Li \and Xiaoming Ma \and Hua Pang ({\large\Letter}) \and Fashen Li \at
              Institute of Applied Magnetics, Key Lab for Magnetism and Magnetic Materials of the Ministry of Education, Lanzhou University, Lanzhou 730000, Gansu, People's Republic of China. \\
              \email{hpang@lzu.edu.cn}
}

\date{Received: date / Accepted: date}

\maketitle

\begin{abstract}
In the present paper, we report a comparison study of SrFe$_2$As$_2$ and SrFeAsF using M\"ossbauer spectroscopy. The temperature dependence of the magnetic hyperfine field is fitted with a modified Bean-Rodbell model. The results give much smaller magnetic moment and magneto-structural coupling effect for SrFeAsF, which may be understood as due to different inter-layer properties of the two compounds.

\end{abstract}

\section{Introduction}
\label{intro}

The surprising discovery of superconductivity by Hideo Hosono and co-workers in fluorine-doped LaFeAsO with transition temperature as high as 26\,K \cite{Kamihara2008} has stimulated worldwide efforts to study this new family of superconductors \cite{Paglione2010}. Among many discovered iron-based superconductors, the 122-type AeFe$_2$As$_2$ (Ae=Ba, Sr, Ca, Eu) and the 1111-type ReFeAsO (Re=La, Ce, Pr, Nd, etc.) families are the most studied materials \cite{Paglione2010}. And it has been shown that magnetism and structural tuning may play important roles in the pairing mechanism of these superconductors \cite{Lumsden2010,LeeCH2008}. There is also an interesting interplay between the crystal structure and magnetism degrees of freedom \cite{Egami2010}. However, a clear understanding of the relationship between magnetism and structural tuning is far from reached.

M\"ossbauer spectroscopy has been proved to be a useful tool to probe the local magnetic field at the iron nucleus of these materials \cite{Blachowski2011,Zhiwei2011}. A comparison study of the 122-type and 1111-type families may yield valuable information about the magneto-structural coupling effect (MSCE). Therefore, in the present work, SrFe$_2$As$_2$ and SrFeAsF compounds are prepared and studied through M\"ossbauer spectroscopy. The result suggests that the MSCE for SrFe$_2$As$_2$ is stronger than that for the SrFeAsF compound, which may be due to the different properties of Sr and SrF inter-layers for SrFe$_2$As$_2$ and SrFeAsF, respectively.

\section{Experimental}
\label{exp}

Polycrystalline SrFe$_2$As$_2$ and SrFeAsF compounds were prepared using conventional solid-state reaction method similar to previous reports \cite{Rotter2008,Han2008}.
Phase purity were checked by X-ray powder diffraction (XRPD) on a Philips X'pert diffractometer with Cu K$_\alpha$ radiation.
The lattice constants were found to be $a$=3.925\,{\AA}, $c$=12.365\,{\AA} for SrFe$_2$As$_2$ and $a$=4.002\,{\AA}, $c$=8.970\,{\AA} for SrFeAsF, which are in good agreement with previously reported values \cite{Rotter2008,Han2008}.
Transmission M\"ossbauer spectra at temperatures between 16\,K and 300\,K were recorded using a conventional constant acceleration spectrometer with a $\gamma$-ray source of 25\,mCi $^{57}$Co in palladium matrix.

\section{Results and Discussion}

\begin{figure}[htp]
  \includegraphics[width=0.8\textwidth]{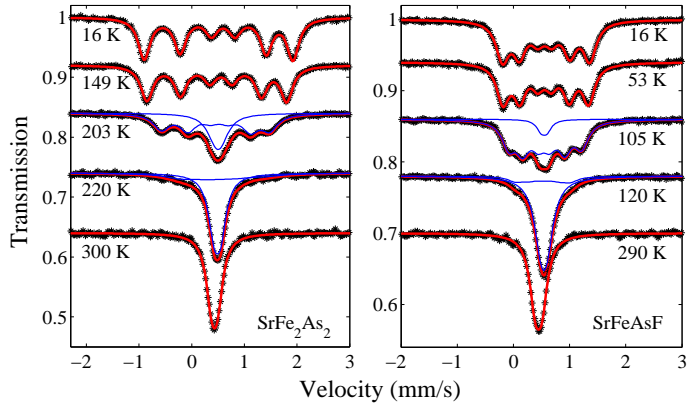}
\caption{M\"ossbauer spectra taken at indicated temperatures of the SrFe$_2$As$_2$ and SrFeAsF compounds.}
\label{fig:Moss}       
\end{figure}

M\"ossbauer spectra (MS) at selected temperatures are shown in Fig. \ref{fig:Moss}. As can be seen, the spectra for both SrFe$_2$As$_2$ and SrFeAsF compounds can be well fitted by only one doublet/(sextet) above/(below) the transition region, indicating that there is no Fe containing impurity phase in our sample, coincidence with the XRPD results. At temperatures in the transition region the MS are fitted with the superposition of a doublet and a sextet. This means that the spin density wave (SDW) develops below or about onset of the magnetism and rapidly reaches quasi-rectangular form upon lowering of the sample temperature \cite{Blachowski2011}.

\begin{figure}[htp]
  \includegraphics[width=0.6\textwidth]{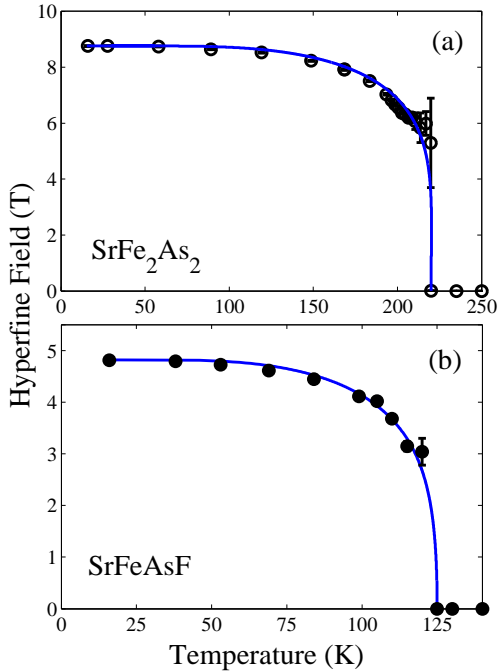}
\caption{Temperature dependence of the hyperfine field, $B_{hf}(T)$, extracted from least-squares fits of the M\"ossbauer spectra. The blue solid lines are theoretical fits to the data (see text).}
\label{fig:Bhf}       
\end{figure}

The variation of the HF with temperature is shown in Fig. \ref{fig:Bhf}. To investigate the MSCE, we fit the temperature dependence of the HF with a modified Bean-Rodbell model. To include MSCE, we rewrite the Gibbs free energy per unit volume as \cite{Zhiwei2011,BeanRodbell1962}
$G =
-\frac{1}{2}Nk_BT_Nm^2 + \frac{1}{2}B\varepsilon^2 + \lambda\varepsilon m^2 - TNk_B[\ln 2 - \frac{1}{2}\ln(1-m^2) - m\tanh^{-1}m],$
where $m$ is the sublattice magnetization, $\varepsilon$ the strain, $T_N$ the N\'{e}el temperature, B the elastic modulus, $k_B$ the Boltzmann constant, and $\lambda$ is the coupling coefficient. Following the same procedure as done by Bean and Rodbell, first minimizing the total energy with respect to $\varepsilon$ and then with respect to $m$ one can obtain the implicit dependence of $m$ on $T$ as \cite{BeanRodbell1962}
$T/T_N = (m/\tanh^{-1}m)(1 + \eta m^2/3),$
where $\eta = 6 \lambda^2/T_NBNk_B$ is a fitting parameter, which involves the magneto-structural coupling coefficient and elastic modulus and controls the order of the magnetic phase transition with $\eta<1$ for a second order phase transition and $\eta>1$ for a first order phase transition. Obviously, a larger $\lambda$ will lead to a larger $\eta$, which will creates a larger energy barrier in the free-energy landscape when $\eta>1$ \cite{Zhiwei2011}. This should leads to a sharper magnetic transition as shown in Fig. \ref{fig:Bhf} for SrFe$_2$As$_2$.

The fitted results are also plotted in Fig. \ref{fig:Bhf} (solid curve). The zero temperature HF is found to be 8.76\,T and 4.82\,T for SrFe$_2$As$_2$ and SrFeAsF, respectively. And $T_N$ is determined to be 220\,K for SrFe$_2$As$_2$ and 125\,K for SrFeAsF, which are in good agreement with reported values \cite{Rotter2008,Han2008}. The structural factor $\eta$ is found to be 1.09 and 0.8 for SrFe$_2$As$_2$ and SrFeAsF, respectively, indicating stronger MSCE for SrFe$_2$As$_2$ than SrFeAsF. The only difference for the two compounds is the non-superconducting layers, namely Sr layer and F-Sr-F layer for SrFe$_2$As$_2$ and SrFeAsF, respectively. From the XRPD results one can obtain that a larger F-Sr-F layer leads to a larger distance along the $c$-direction between the FeAs plane for the SrFeAsF, which should reduce the effective interactions between the FeAs layers. Moreover, due to different properties of Sr and F-Sr-F inter-layers for SrFe$_2$As$_2$ and SrFeAsF, the crystal field for the two compounds should be different. This could lead to a completely different electronic structure for the two compounds, which may lead to the observed different magnetic moments and MSCE.

\section{Conclusions}
In summary, our M\"ossbauer results show that SrFe$_2$As$_2$ have a much larger magnetic moment than SrFeAsF. And the antiferromagnetic phase transition is sharper for SrFe$_2$As$_2$, indicating a stronger magneto-structural coupling effect than that in SrFeAsF, which might due to different inter-layer properties of the two compounds.



\end{document}